\documentclass[twocolumn]{svjour3} 
\smartqed  
\usepackage[caption=false,font=footnotesize]{subfig}
\usepackage{latexsym}
\usepackage{etoolbox}
\usepackage{url}
\usepackage{booktabs}
\usepackage{algorithmic}
\usepackage{array}
\usepackage{graphicx}
\usepackage{textcomp}
\usepackage{xspace}
\usepackage{xcolor}
\usepackage[version-1-compatibility]{siunitx}
\usepackage{multirow}

\newcommand{\DAVIDE}{D.A.V.I.D.E.\xspace}
\newcommand{\DIG}{DiG\xspace}
\newcommand{\quotes}[1]{``#1''}

\begin{document}

\title{\DIG: Enabling Out-of-Band Scalable High-Resolution Monitoring for Data-Center Analytics, Automation and Control (Extended)\\
}

\titlerunning{DiG} 

\author{Antonio Libri \and Andrea Bartolini \and Luca Benini}


\institute{Antonio Libri \at
              D-ITET, ETH Zurich, Zurich, Switzerland\\
              \email{a.libri@iis.ee.ethz.ch}
           \and
           Andrea Bartolini \at
              DEI, University of Bologna, Bologna, Italy\\
              \email{a.bartolini@unibo.it}
           \and
           Luca Benini \at
              D-ITET, ETH Zurich, Zurich, Switzerland\\
              DEI, University of Bologna, Bologna, Italy\\
              \email{lbenini@iis.ee.ethz.ch, luca.benini@unibo.it}
}

\date{Received: date / Accepted: date}

\maketitle

\begin{abstract}
\sloppypar
Data centers are increasing in size and complexity, and we need scalable approaches to support their automated analysis and control. Performance counters and power consumption are their key \quotes{vital signs}. State-of-the-Art (SoA) monitoring systems provide built-in tools to collect performance measurements, and custom solutions to get insight on their power consumption. However, with the increase in measurement resolution (in time and space) and the ensuing huge amount of measurement data to handle, new challenges arise, such as bottlenecks on the network bandwidth, storage and software overhead on the monitoring units. To face these challenges we propose a novel monitoring platform for data centers, which enables real-time high-resolution profiling (\textit{i.e.}, all available performance counters and the entire signal bandwidth of the power consumption at the plug - sampling up to \SI{20}{\micro\s} - with an error below \SI{1}{\percent}) and analytics, both at the edge (node-level analysis) and on a centralized unit (cluster-level analysis). The monitoring infrastructure is completely out-of-band, scalable, technology agnostic and low cost, and it is already installed in a SoA high-performance compute cluster (\textit{i.e.}, \DAVIDE~- 18\textsuperscript{th}~in Green500 November 2017).

\end{abstract}

\section{Introduction}\label{sec:intro}

Data centers and High Performance Computing (HPC) systems are becoming increasingly complex and the need for novel methods to support their automation, analytics and control is garnering considerable attention~\cite{liu2018}. In this direction, industry and academia researchers are pushing toward the use of Artificial Intelligence (AI) and Machine Learning (ML) techniques to address non-trivial challenges such as efficient management of computational\,/\,infrastructure resources, detection of anomalies and failures, and predictive maintenance. As an example, Duplyakin et al.~\cite{dup2016} have shown how to get high-confidence predictions of the time-to-completion and energy consumption of scientific applications (which can help for a more efficient usage of the resources) via Active Learning techniques applied to regression problems. Other examples are based on unsupervised learning, such as~\cite{ahmad2017} which introduces methods for real-time anomaly detection on streaming data (useful for an early warning about problems in the system and the hosted applications), and~\cite{tang2014} that shows a way to detect malware using hardware features.

All these techniques exploit low-level monitoring of the hardware of the data-center infrastructure (\textit{i.e.}, application and system performance, and related power and energy consumption). In particular, depending on the target use-case, some features can reveal more information than others: a highly flexible monitoring has to collect as many metrics as possible. On the other hand, this implies to face three main bottlenecks related to the large amount of monitoring data produced: (i)~overhead on the network's bandwidth, (ii)~overhead on the data storage capacity (to save measurements for post-processing analysis) and (iii)~overhead on the software tools that have to handle the measurements (in real-time and offline). 

\begin{figure}[ht]
  \centering
  \includegraphics[width=8cm]{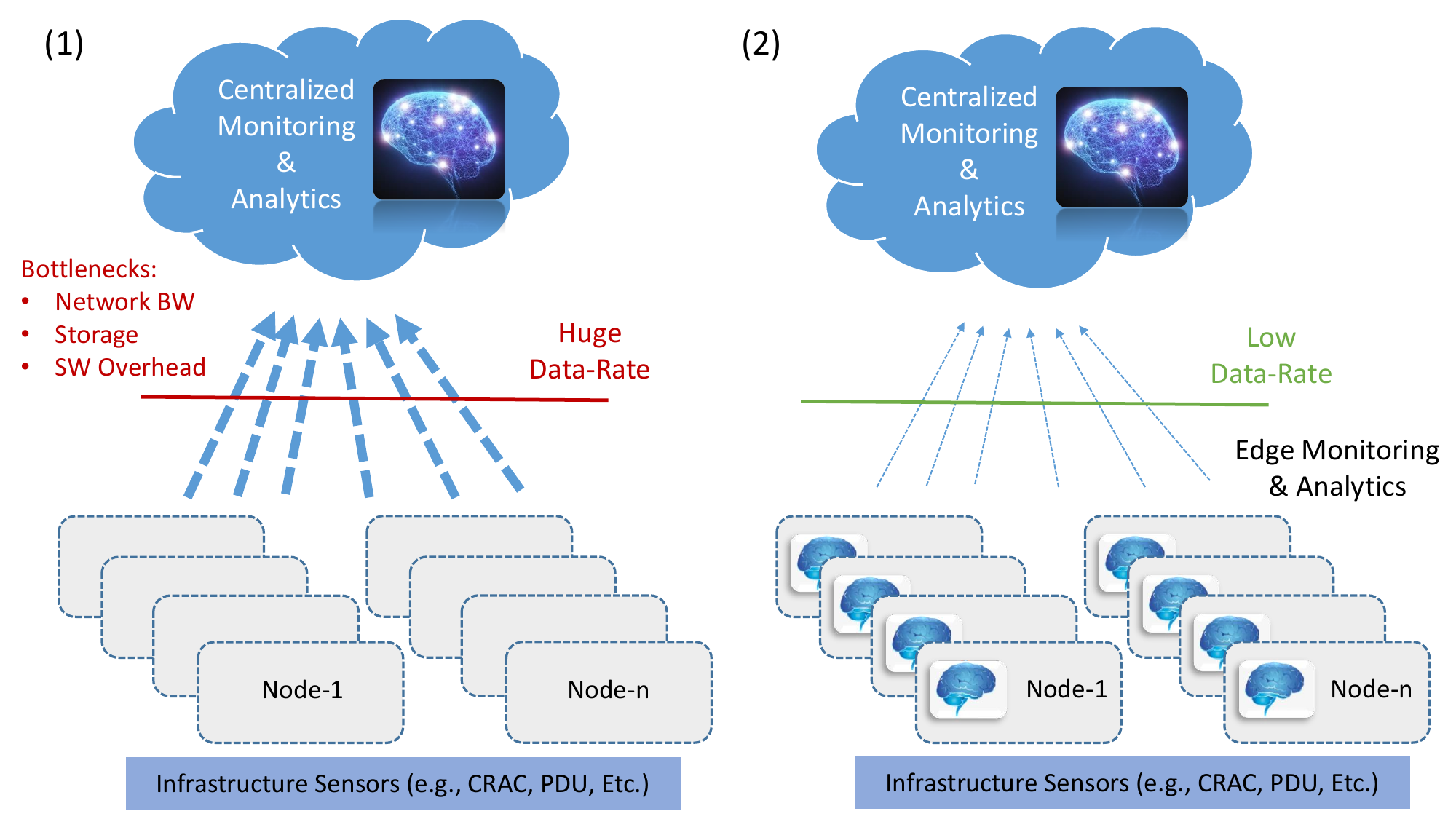}
  \caption{Data-center monitoring design and bottlenecks.}
  \label{fig:datacenter_mon}
\end{figure}

This is depicted in Figure~\ref{fig:datacenter_mon} (left), which shows that the node dedicated to the monitoring software stack has the complete view of the status of the cluster (and thus can exploit measurements for ML analysis), but has to deal with the above mentioned bottlenecks. To give an example, Ilsche et al. developed a high resolution power monitoring system (\textit{i.e.}, HAEC~\cite{Ilsche2018}) that supports a sampling rate of \SI{500}{\kilo S/\s} (kilo Samples per second) on 4 custom sensors. In general, high resolution power monitoring instrumentation is the current trend for both industrial and academia HPC facilities and data centers~\cite{Ilsche2018,HDEEM,PowerInsight}. This is useful to appreciate the power consumption of application phases, but of course the finer the granularity the more difficult is to scale to a large number of nodes in a machine. For instance, instrumenting with HAEC the supercomputer \textit{Sunway TaihuLight} - 2\textsuperscript{nd} in Top500 of June 2018 and that includes around 41 thousand computing nodes~\cite{Fu2016} - would require a data collection network bandwidth of around \SI{82}{\giga S/\s }, with obvious overheads on software and storage to handle it.

An intuitive solution is to bring some of the \quotes{monitoring intelligence} to the edge and codesign the monitoring infrastructure to leverage data analysis between distributed monitoring agents and a centralized unit. This is represented in Figure~\ref{fig:datacenter_mon} (right), which shows distributed smart monitoring agents that can carry out real-time analysis per node (\textit{e.g.}, feature extraction, ML inference, etc.) and share information with the centralized monitoring at a much lower rate (\textit{e.g.}, detection of an anomaly in a node, plus other measurements at a lower rate needed for cluster level analysis). In this distributed architecture, each monitoring agent has the complete knowledge of the status of its node, while the centralized monitoring unit has the complete view of the cluster, thus can carry out analysis at a higher level.

\sloppypar
Current State-of-the-Art (SoA) monitoring solutions allow to collect measurements in-band and out-of-band by means of built-in tools (\textit{e.g.}, Amester~\cite{rosedahl2017} or RAPL~\cite{rapl_Khan2018}, which expose hardware performance counters) or custom sensors (\textit{e.g.}, HDEEM~\cite{HDEEM} or HAEC~\cite{Ilsche2018}, which provide fine grain power measurements), where the benefit of the out-of-band solution is no overhead on the computing resources. However, to the best of our knowledge, there is not yet a monitoring infrastructure for compute clusters and high performance machines that provides a flexible way to analyze all possible features (\textit{i.e.}, all available hardware performance counters and the entire signal bandwidth of the node's power consumption). This paper, which builds upon and extends our previous publication~\cite{DiG_DAAC18}, focuses on a novel scalable and high resolution monitoring infrastructure for data centers and HPC systems. The system is completely out-of-band and provides a highly flexible environment to work both at the edge and at cluster-level for data center analytics, automation and control.

\textit{Contributions of the work:} 
\begin{enumerate}
    \item design of an out-of-band monitoring infrastructure - we named it \DIG (\textit{i.e.}, Dwarf in a Giant) - that exploits edge monitoring agents and centralized cluster-level monitoring for data centers analytics, automation and control. The platform design provides a highly flexible environment to tackle different challenges. We designed a custom power sensor at the plug to monitor the power consumption at high resolution, covering the entire signal bandwidth (sampling up to \SI{20}{\micro \s}) with a measurements precision below \SI{1}{\percent} ($\sigma$) (therefore also suitable for the most rigorous power measurement requirements to benchmark a computing system in Top500~\cite{EEHPCWG}). The system allows to interface with existing out-of-band telemetry (\textit{e.g.}, Amester~\cite{rosedahl2017}, IPMI~\cite{IPMI}), but also with in-band built-in tools if required (\textit{e.g.}, RAPL~\cite{rapl_Khan2018}). All the measurements are synchronized at sub-microseconds precision to obtain a detailed picture over time of the nodes and cluster state. We adopted a scalable and lightweight interface to the centralized monitoring (\textit{i.e.}, MQTT~\cite{mqtt_ibm}) to support large-scale computing centers. The monitoring infrastructure is technology agnostic (\textit{i.e.}, already tested on different architectures, such as Intel, ARM and IBM) and low cost (\textit{i.e.}, the custom power sensor does not require any motherboard redesign).
    \item we report (i) the performance of the monitoring agents (\textit{i.e.}, measurements granularity, precision, synchronization, software overhead and scalability), along with (ii) an extensive campaign of ML inference benchmarks running on the dedicated embedded computers and based on deep Residual Networks (a.k.a. ResNets~\cite{ResNet2016}), to obtain an assessment of their real-time inference capabilities, and (iii) an extensive set of tests based on frequency-domain analysis to show the capability of the high resolution monitoring to unveil high-frequency components directly related to the computation activity and including also two use-cases that can be used for anomaly detection.
    \item we validated and calibrated our high-resolution power measurements, and provide detailed information on accuracy and precision (best in class w.r.t SoA data centers monitoring systems); moreover, we integrated the monitoring infrastructure in a SoA HPC cluster (\textit{i.e.}, \DAVIDE~\cite{DAVIDE}~- 18\textsuperscript{th} in Green500 November 2017) that is already in production and available to the users community since more than one year.
    \item We provide detailed information of both hardware and software architectures for the whole monitoring infrastructure (Sections~\ref{sub:sensing}, \ref{sub:embedd}, \ref{sub:front-end} describe the edge infrastructure, while the cluster-level software - namely ExaMon - can be downloaded from~\cite{ExaMon}).
\end{enumerate}

\textit{Outline:} Section~\ref{sec:dig} presents the monitoring architecture. Its performance is analyzed in Section~\ref{sec:results}, together with several case studies of frequency-domain analysis on the high-resolution power measurements. We report related works in Section~\ref{sec:rw} and conclude the paper in Section~\ref{sec:end}.

\section{Monitoring System Architecture}\label{sec:dig}

One of the main challenges we faced during the monitoring system design was to make it suitable for different hardware architectures and low cost. With this goal, we targeted only what is missing on today's built-in monitoring solutions~\cite{Ilsche2018}: a custom power sensor that allows high resolution monitoring. We placed it at the node power source to completely avoid motherboard re-design or modification. We then interfaced it with a dedicated low-cost embedded computer (one per node) that is suitable for monitoring applications.

A second challenge was to make the system highly flexible in terms of monitoring capabilities. With this goal, we interfaced the embedded computer with built-in tools to get per-component monitoring (\textit{i.e.}, hardware performance counters) and have the complete knowledge of the status of the node. This information, along with the high resolution power monitoring, can reveal not only insights on application behavior but also patterns on performance\,/\,failures of components (\textit{e.g.}, fans, HDDs).

\begin{figure}[ht]
  \centering
  \includegraphics[width=8cm]{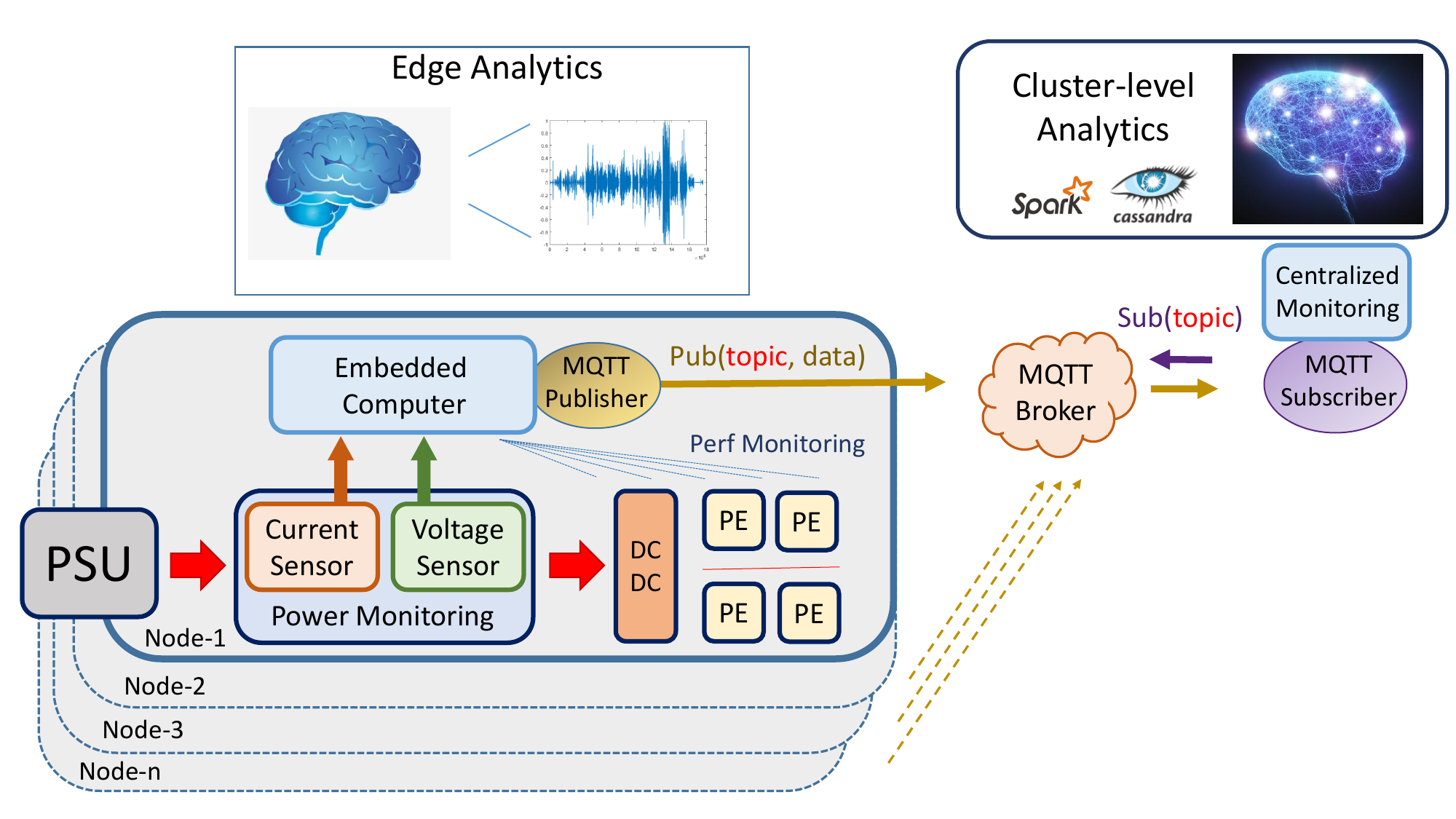}
  \caption{Sketch of the monitoring system architecture.}
  \label{fig:block_diagram}
\end{figure}

Finally, we exploited a scalable and lightweight interface (\textit{i.e.}, MQTT) to send information to a centralized monitoring unit and perform cluster-level analytics. Figure~\ref{fig:block_diagram} shows the main components of the monitoring system that will be described in this section. 

\subsection{High-Resolution Power Sensing}\label{sub:sensing}

\begin{figure}[t]
  \centering
  \includegraphics[width=8.3cm]{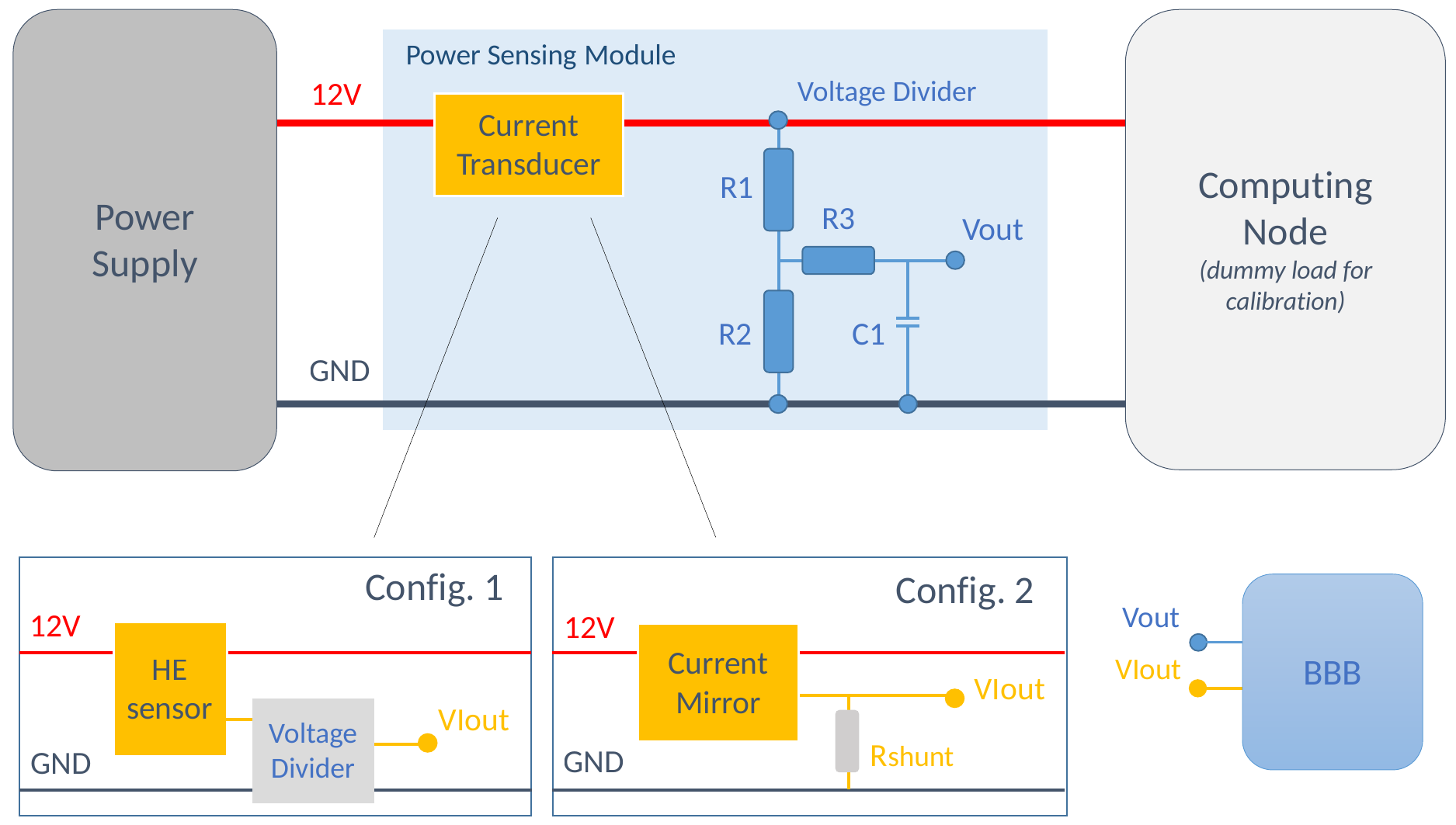}
  \caption{High-Resolution power sensing module.}
  \label{fig:schematic}
\end{figure}

To provide high resolution measurements of the nodes' power consumption we placed a power sensing module between the Power Supply Unit (PSU) and the DC-DC converters that provide power for all the processing elements (PE)\,/\,electrical components within the node. Figure~\ref{fig:schematic} shows the schematic of the power sensing module. We use a voltage divider based on high precision resistors to measure the voltage and a current transducer to measure the current. Their outputs are then connected to the ADC integrated in the embedded monitoring board, via first-order low-pass filters needed to counter aliasing effects. Indeed, due to the high operating frequencies of data centers\,/\,HPC nodes, the power consumption is highly dynamic, and therefore an anti-aliasing filter is required. We have chosen a voltage divider as it provides a simple but effective solution to properly scale the voltage in input to the ADC without any additional hardware (\textit{e.g.}, an isolated power supply would be needed if using active components). 

For the current transducer, we tested two configurations: one based on a Hall Effect (HE) sensor and one based on a current mirror and shunt resistor. Thanks to the high output linearity, both solutions obtain satisfactory results. We tested the first configuration with Intel Xeon E5-2600 (Haswell) and ARM Cavium TunderX architectures, while the second one on a cluster based on OpenPOWER IBM Power8~\cite{DAVIDE}. In particular, the HE sensor is the \textit{Allegro MicroSystems ACS770}~\cite{allegro}. It can measure currents in the range \SIrange[range-phrase = --]{0}{100}{\ampere} with low-intrusiveness, good linearity and high precision sensitivity (\SI{40}{\mV/\ampere}). It has a typical bandwidth of \SI{120}{\kHz} and an internal conductor resistance of \SI{100}{\micro \ohm}, which implies a negligible power loss. All these features make it suitable for integration with standard PSUs used on Intel and ARM architectures.

For the configuration based on current mirror and shunt resistor, we exploited the same current mirror already used by the Baseboard Management Controller (BMC) on IBM Power8 to measure at a coarse grain the total node power consumption. This provides a twofold benefit: measuring currents in a wider range (\SIrange[range-phrase = --]{0}{250}{\ampere}) and avoiding extra cost for current sensing components. Finally, to avoid a possible measurement accuracy degradation due to heating effects on the resistors of the voltage dividers, we have chosen resistors with equal Temperature Coefficient of Resistance (TCR), and placed them close to each other (similar temperature on both) and far from external sources of heating (to avoid uneven heating effects).

\subsection{Embedded Monitoring and Edge Analytics}\label{sub:embedd}

We selected a Beaglebone Black (BBB)~\cite{bbb} as embedded computing board as it provides several interesting features off-the-shelf: (i)~it includes a 12-bit Successive Approximation Register (SAR) ADC needed for the power-sensing module, (ii)~hardware support for Precision Time Protocol (PTP) which allows sub-microsecond measurements synchronization~\cite{my_hpcs2016,myAndare18}, and (iii)~an ARM Cortex-A8 processor with NEON technology, useful for DSP processing and edge ML inference (\textit{e.g.}, by leveraging the ARM NN SDK~\cite{ArmNN}, which enables efficient translation of existing neural network frameworks, such as TensorFlow, to ARM Cortex-A CPUs). Moreover, the Sitara AM335x chip used in BBB includes two programmable real-time units (PRUs) useful for real-time acquisition and extra processing on-board.

It should be noted that standard computing servers already integrate an embedded system used for real-time monitoring and management, namely the BMC.
This is usually a closed platform with no access to the firmware, but thanks to the recent OpenBMC project~\cite{rosedahl2017} few vendors started to release it open-source. We decided to do not use the BMC for this purpose as (i)~we needed to add extra hardware to integrate an ADC and interface it with the custom power sensor (the BBB already includes an ADC), (ii)~it does not provide hardware support for PTP, and (iii)~it is based on an old ARM processor family (\textit{i.e.}, ARM11~\cite{AST2500}) which is not a good choice for edge analytics. Moreover, (iv) it is a critical component to ensure safe working conditions of the nodes, thus it is more convenient to do not overwhelm its limited processing resources with our monitoring\,/\,edge-analytics software stack.

\begin{figure}[t]
  \centering
  \includegraphics[width=8cm]{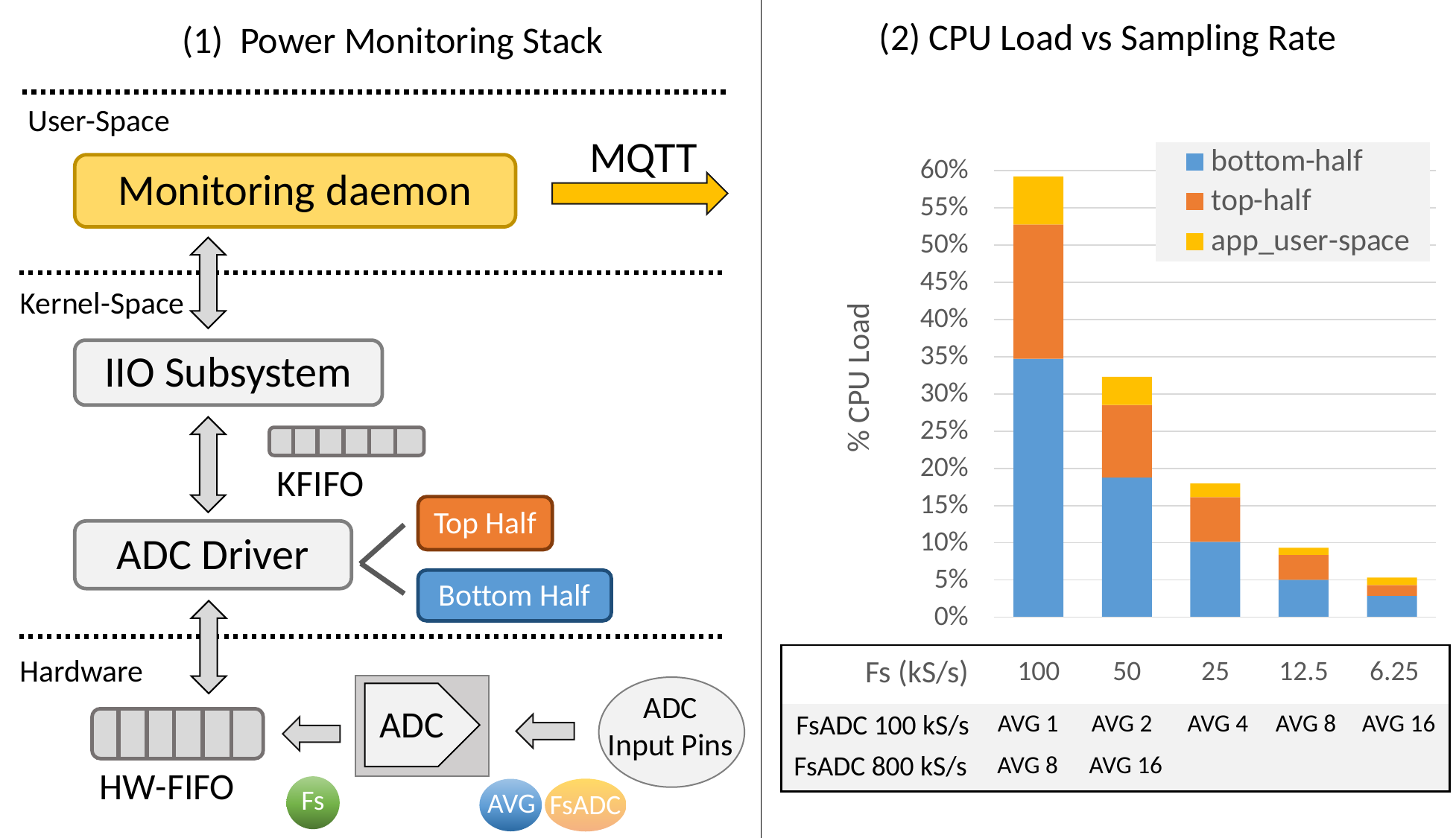}
  \caption{Embedded power monitoring stack (left) and software overhead (right).}
  \label{fig:adc}
\end{figure}

Figure~\ref{fig:adc} reports the embedded monitoring stack design. The bottom layer represents the ADC hardware module. We exploit the ADC \textit{continuous sampling mode} to continuously sample the two input channels (\textit{i.e.}, current and voltage), average and store them in a hardware FIFO that is managed by a kernel driver. By tuning the ADC sampling frequency (FsADC) and the hardware averaging (AVG), it is possible to set the frequency (Fs) at which samples enter the software layers.

When the hardware FIFO reaches a watermark on the number of samples acquired (we set it to 16), an interrupt is raised and the samples are flushed into the main memory (kernel FIFO) by the ADC driver. In particular, the ADC driver involves two routines (\textit{IRQ handlers}): the \textit{top half}, which monitors the watermark on the hardware FIFO, and the \textit{bottom half} that is used to flush the samples into the kernel FIFO.

Finally, the power measurements are exposed via the Industrial I\,/\,O (IIO) Subsystem API to a user-space monitoring daemon, which is in charge of converting data from integer to Watt and associate them with a timestamp. This daemon also collects node's performance measurements from hardware performance counters via built-in tools (\textit{e.g.}, IPMI~\cite{IPMI}, Amester~\cite{rosedahl2017} and RAPL~\cite{rapl_Khan2018}). In this way, we can perform edge analytics on a target use-case (\textit{e.g.}, ML inference for anomalies detection) and send the results, together with the power and performance measurements at a lower rate, to the centralized monitoring unit for cluster-level analytics.

It should be noted that by using the continuous sampling mode, the hardware guarantees a negligible uncertainty on the acquisition time of consecutive samples, ensuring correctness of the energy computation at a fine granularity~\cite{HDEEM}. Moreover, to make negligible both overhead and uncertainty introduced by the timestamp's function call, we only generate timestamps at each flush of the kernel FIFO, and not for every sample (timestamps for every sample are then derived accordingly to the number of samples acquired).

\subsection{Centralized Monitoring Unit and Cluster-Level Analytics}\label{sub:front-end}

We exploit centralized monitoring, based on the open-source ExaMon monitoring platform~\cite{ExaMon,mqtt_beneventi}, to carry out cluster-level analytics with data coming from multiple nodes. To send data from the distributed monitoring agents to the centralized monitoring unit, we adopted MQTT~\cite{mqtt_ibm}, which is a robust, lightweight and scalable protocol, already used for large-scale systems both in industry and academia (\textit{e.g.}, Amazon, Facebook,~\cite{mqtt_ibm,mqtt_beneventi}). Figure~\ref{fig:block_diagram} outlines its publish\,/\,subscribe communication model, where the publishers (running in the embedded computers) send measurements to a broker, along with a topic that corresponds to the monitored metric (\textit{e.g.}, power consumption). The broker resides in the centralized monitoring unit together with the subscriber. The latter is used to filter and collect the data that is interested in, and expose them to a Big Data engine, namely Apache Spark~\cite{spark2016}. The measurements are also stored in a scalable database - Apache Cassandra~\cite{cassandra} - enabling ML analytics both in streaming and batch mode.

\section{Experimental Results}\label{sec:results}

This section starts by reporting the performance of the monitoring agents, along with the validation of the high-resolution power measurements from the point of view of accuracy and precision. It then presents a set of monitoring use cases based on Fourier analysis to show the capability of the high-resolution monitoring in revealing fine grain computation activity. Finally, we report a campaign of ML inference benchmarks, based on ResNets running on the embedded monitoring platform, to show an example of the capability of \DIG on carrying out edge ML analytics.

\subsection{Monitoring Agents Performance}\label{sub:performance}

\textit{Performance Measurements:} To provide completely out-of-band monitoring, we integrated our infrastructure in a SoA OpenPOWER computing cluster, namely \DAVIDE~\cite{DAVIDE,BigDaw}, that consists of 45 nodes (3 racks with 15 nodes each) based on IBM Power8. In this system we take advantage of its out-of-band telemetry and collect via Amester through the On Chip Controller (OCC)~\cite{rosedahl2017} 242 metrics per-component every \SI{10}{\s} (\textit{e.g.}, the performance of Core, Cache, FAN, etc.), and via IPMI 89 metrics per-component every \SI{5}{\s}. All these measurements are sent to the centralized monitoring for cluster-level analytics, but can also be exploited on-board for real-time edge analysis on a target use-case (\textit{e.g.}, detection of anomalies).

\textit{Power Measurements:} To cover the full power consumption bandwidth at the node plug (\textit{i.e.}, tens of microseconds, observed with a professional oscilloscope - \textit{Keysight DS0X3054T} - attached to the power sensor) and at the same time to avoid overwhelming the CPU with the data acquisition routine, we tested several sampling rates (Fs) that we report in Figure~\ref{fig:adc}.2. In particular, the best trade-off corresponds to a sampling rate of \SI{800}{\kilo S/\s} per channel and hardware averaging every 16 samples. This is equivalent to \SI{50}{\kilo S/\s} (\textit{i.e.}, \SI{20}{\micro \s}) and allows to obtain several benefits: (i) to cover the entire signal bandwidth, (ii) to keep the CPU load below \SI{35}{\percent} and (iii) to reach a precision below \SI{1}{\percent} ($\sigma$) of uncertainty (a.k.a. oversampling and averaging method~\cite{BitRes}). It is noteworthy that this precision makes our system suitable for the most rigorous requirement needed to benchmark an HPC system in Top500~\cite{EEHPCWG}. 

Finally, we send the measurements to the centralized monitoring for cluster-level analytics at the rates of \SI{1}{\s} and \SI{1}{\milli \s}, while measurements at higher resolution can be analyzed directly at the edge.

\textit{Software overhead:} The entire BBB CPU load of the monitoring software stack (power and performance) is below \SI{46}{\percent}. In particular, performance monitoring requires roughly \SI{11}{\percent} and the power monitoring around \SI{35}{\percent}. We have run also some benchmarks to evaluate the computing capabilities of the embedded platform: we can perform (i)~real-time Power Spectral Density (PSD) analysis of the high resolution power measurements (\textit{e.g.}, useful for feature extraction~\cite{DNN_FFT_2018}), in a time window of around \SI{40}{\milli\s} with roughly \SI{7}{\percent} of CPU usage, and (ii)~ML inference via TensorFlow on these PSDs, implementing ResNet with 8 layers and channels $\{8,8,16,32\}$ and respecting a real-time constraint of \SI{30}{\milli \s} per spectrogram (a detailed analysis can be found in Section~\ref{sub:edge_ML}). Moreover, it must be noted that we did not use any optimization to run ML inference (\textit{e.g.}, ARM NN SDK~\cite{ArmNN} or TensorFlow Light~\cite{TF_Light}), which means these results can be further improved.

\textit{Synchronization:} To ensure accurate and precise timestamping of the measurements, we exploit PTP hardware obtaining sub-microsecond synchronization (\textit{i.e.}, smaller than the sampling period) across multiple nodes and embedded monitoring devices. Moreover, results in~\cite{myAndare18} show that with a proper setting of NTP, it is possible to reach synchronization with an uncertainty of a few microseconds in today's data centers and HPC clusters.

\textit{Scalability:} To benefit from a scalable interface to the centralized monitoring unit, we exploit Mosquitto~\cite{mqtt_beneventi} which is a Linux implementation of MQTT that consists of a single thread process. Our tests show that Mosquitto broker (\textit{i.e.}, the bottleneck of the network) running in an Intel Xeon E5-2600 (Haswell) can handle up to 16 publishers that send data every millisecond using just \SI{30}{\percent} of a core, and of course, it is possible to increase the number of brokers if needed. In our current configuration of the monitoring system integrated in the \DAVIDE HPC machine, we use one broker for all performance counters and three brokers (one per rack) for the power measurements, with no particular issue for the users\,/\,system admins of the data center since November 2017. Moreover, we tested MQTT with a similar configuration on all 516 computing nodes of GALILEO at CINECA, Italy (Intel Xeon E5-2630v3 processors), proving that this interface is suitable for even larger scale systems.

\subsection{Power Measurements Validation}\label{sub:validation}

\begin{figure}[ht]
  \centering
  \includegraphics[width=8.2cm]{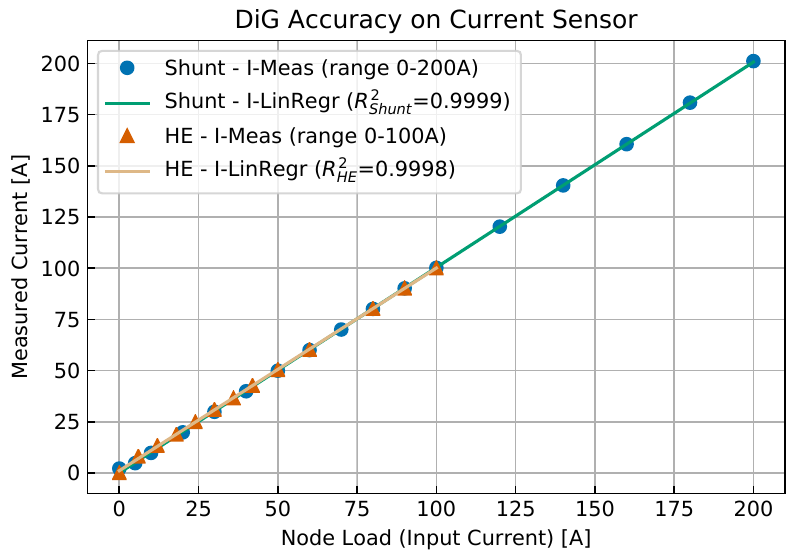}
  \caption{Accuracy of the \DIG current sensors.}
  \label{fig:dig_acc}
\end{figure}

To verify the accuracy and precision of the high resolution power measurements, we attached the \DIG power sensing module to a dummy load (depicted in Figure~\ref{fig:schematic}) and calibrated the current and voltage sensors against a high-precision reference multimeter. In this way, we can also determine the conversion factors (offset and gain) for both voltage and current needed for computing power consumption in Watt. Figure~\ref{fig:dig_acc} reports the results of the current measurements after calibration for both configurations, namely HE Sensor and current mirror plus shunt resistor. In particular, the x-axis corresponds to the different input loads that we applied to the power sensing module accordingly to the allowed current range (\textit{i.e.}, \SIrange[range-phrase = --]{0}{100}{\A} for the HE Sensor and \SIrange[range-phrase = --]{0}{200}{\A} for the shunt resistor), while the y-axis reports the current measured by \DIG (dots) and a linear regression on the measurements (straight line). As can be seen from the plot, all the \DIG measurements well match the input load, so the curve is linear across the full range (\textit{i.e.}, \textit{coefficient of determination} $R^2_{Shunt} = 0.9999$ and $R^2_{HE} = 0.9997$), except around zero, which is anyway not a problem as compute nodes never work in this low range of currents. We carried out the same procedure for the voltage measurements and obtained similar results.

After calibrating \DIG, we can evaluate the precision on the power measurements (\textit{i.e.}, standard deviation - $\sigma$ - and coefficient of Variation - CV). With this goal, we can start by quantifying the precision of each ADC channel (current and voltage) independently, and use the propagation of uncertainty theorem for computing the uncertainty of the resulting power~\cite{uncertainty}. Indeed, given the measured current and voltage with uncertainties, $I \pm \sigma_I$ and $V \pm \sigma_V$ (where $I$ and $V$ correspond to the measured average value), under the assumption they are not correlated, the uncertainty on the power measurements is:

\begin{equation}
    \sigma_P \approx \sqrt{I^2 \sigma_V^2 + V^2 \sigma_I^2}
\end{equation}

Table~\ref{tab:precision} reports the resulting precision at \SI{50}{\kilo S/\s} for three different server operating conditions: idle, medium load and maximum load (\textit{e.g.}, to give an idea in an Intel Xeon E5-2600 these conditions corresponds to roughly \SI{180}{\W}, \SI{600}{\W} and \SI{1200}{\W}, respectively). The precision for around \SI{68.3}{\percent} of the power measurements ($\sigma$) is bounded between \SIrange[range-phrase = --]{1.73}{3.96}{\W}, for the minimum (idle) and maximum load, respectively, and increases of a factor of 3 when considering \SI{99.7}{\percent} of the samples (\SIrange[range-phrase = --]{5.2}{11.88}{\W} for $3 \sigma$). Of course, the CV follows the opposite trend: it decreases when the power consumption increases (from \SI{0.96}{\percent} in idle to \SI{0.33}{\percent} for maximum workload).

The table reports also the precision when sampling at lower rates - by applying a software average -, namely \SI{25}{\kilo S/\s}, \SI{1}{\kilo S/\s} and \SI{1}{S/\s}. As can be seen, $\sigma$ drastically improves to a few watt precision already at \SI{25}{\kilo S/\s} (even at the maximum load). With the goal to dynamically increase the \DIG precision on the power measurements when required, the monitoring daemon can be set to dynamically switch to a lower sampling rate (\textit{e.g.}, to \SI{25}{\kilo S/\s}) by averaging in software the power samples if it is monitoring low currents for a certain time period. Thanks to this trade-off we can always keep the monitoring precision below a pre-set threshold (up to sub-watt precision), which makes \DIG suitable to be used in production environments as a high-precision HPC energy monitoring solution.

\begin{table}[t]
\centering
\caption{\DIG Precision based on dynamic software average.}
\label{tab:precision}
\small\addtolength{\tabcolsep}{-3pt}
\begin{tabular}{>{}l>{}l>{}l>{}l}
 & Idle & Mid-Load & Max Load  \\
 & $\sigma$ (CV) & $\sigma$ (CV) & $\sigma$ (CV)  \\
\noalign{\smallskip}\hline\noalign{\smallskip}
\SI{50}{\kilo S/\s} & \SI{1.73}{\W} (0.96\%) & \SI{2.58}{\W} (0.43\%) & \SI{3.96}{\W} (0.33\%) \\
\SI{25}{\kilo S/\s} & \SI{0.5}{\W} (0.28\%) & \SI{1.26}{\W} (0.21\%) & \SI{2.28}{\W} (0.19\%) \\
\SI{1}{\kilo S/\s} & \SI{0.47}{\W} (0.26\%) & \SI{1.14}{\W} (0.19\%) & \SI{2.16}{\W} (0.18\%) \\
\SI{1}{S/\s} & \SI{0.32}{\W} (0.18\%) & \SI{1.02}{\W} (0.17\%) & \SI{2.04}{\W} (0.17\%) \\
\noalign{\smallskip}\hline
\end{tabular}
\end{table}

\subsection{Feature Extraction Benchmarking}\label{sub:dig_bench}

This section aims at showing the capability of the high resolution monitoring in unveiling high-frequency components directly related to the computation activity. We exploit Fourier analysis as an example of feature extraction technique for time series that is suitable for deep learning algorithms (\textit{e.g.}, Deep Neural Networks - DNNs~\cite{DNN_FFT_2018}). Future works can extend this real-time analysis targeting specific use-cases (\textit{e.g.}, anomaly detection in workloads) and (i)~exploit it, together with performance counters, as input data for DNN models running inference in the monitoring agents~\cite{DNN_FFT_2018}; or (ii)~just send it with lossy compression algorithms (data are sparse, as shown by the following tests) to the centralized monitoring for cluster-level analytics. We note that due to the limitations of SoA power monitoring support for computing nodes, up until now this kind of analysis could not be performed in production data centers. 

\begin{figure}[ht]
  \centering
  \includegraphics[width=8cm]{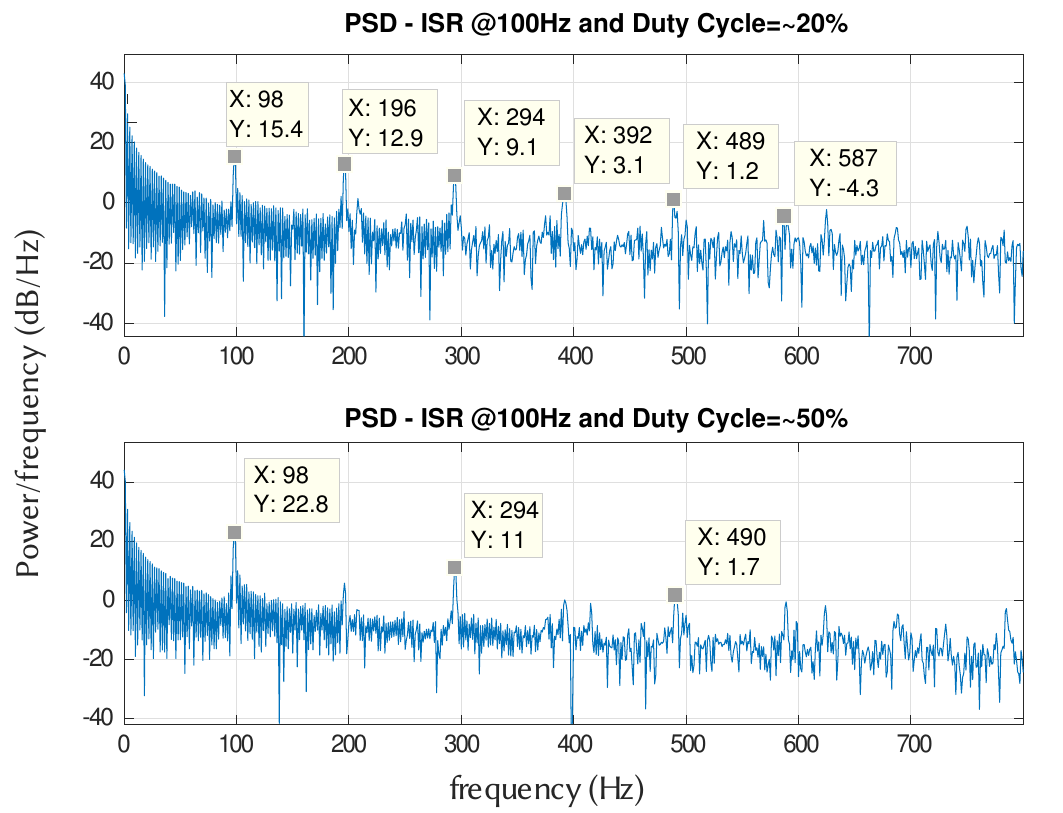}
  \caption{PSD of an ISR at \SI{100}{\Hz} with different sets of instructions.}
  \label{fig:ctrlpulse}
\end{figure}

We start the evaluation with a synthetic benchmark on the computing node, consisting of an Interrupt Service Routine (ISR) that we run every $\sim$\SI{10}{\milli\s} (roughly \SI{100}{\Hz}) with different sets of instructions. In particular, the first set of instructions corresponds to a duty cycle of \SI{20}{\percent} in power consumption (\textit{i.e.}, \SI{2}{\milli\s} of workload and \SI{8}{\milli\s} of sleep), while the second set to \SI{50}{\percent}. Figure~\ref{fig:ctrlpulse} shows the PSD for the two cases, computed in a time window of \SI{40}{\milli \s}. According to Fourier analysis, the set of instructions with duty cycle \SI{20}{\percent} (top) shows the fundamental at around \SI{100}{\Hz} plus all its harmonics, while the one at \SI{50}{\percent} (bottom) correctly reports only the fundamental and the odd harmonics (even harmonics are not completely null due to the not exact \SI{50}{\percent} duty cycle). This example shows that our monitoring can really capture spectral properties of different workloads in execution.

\begin{figure*}[ht]
  \centering
  \includegraphics[width=0.77\textwidth]{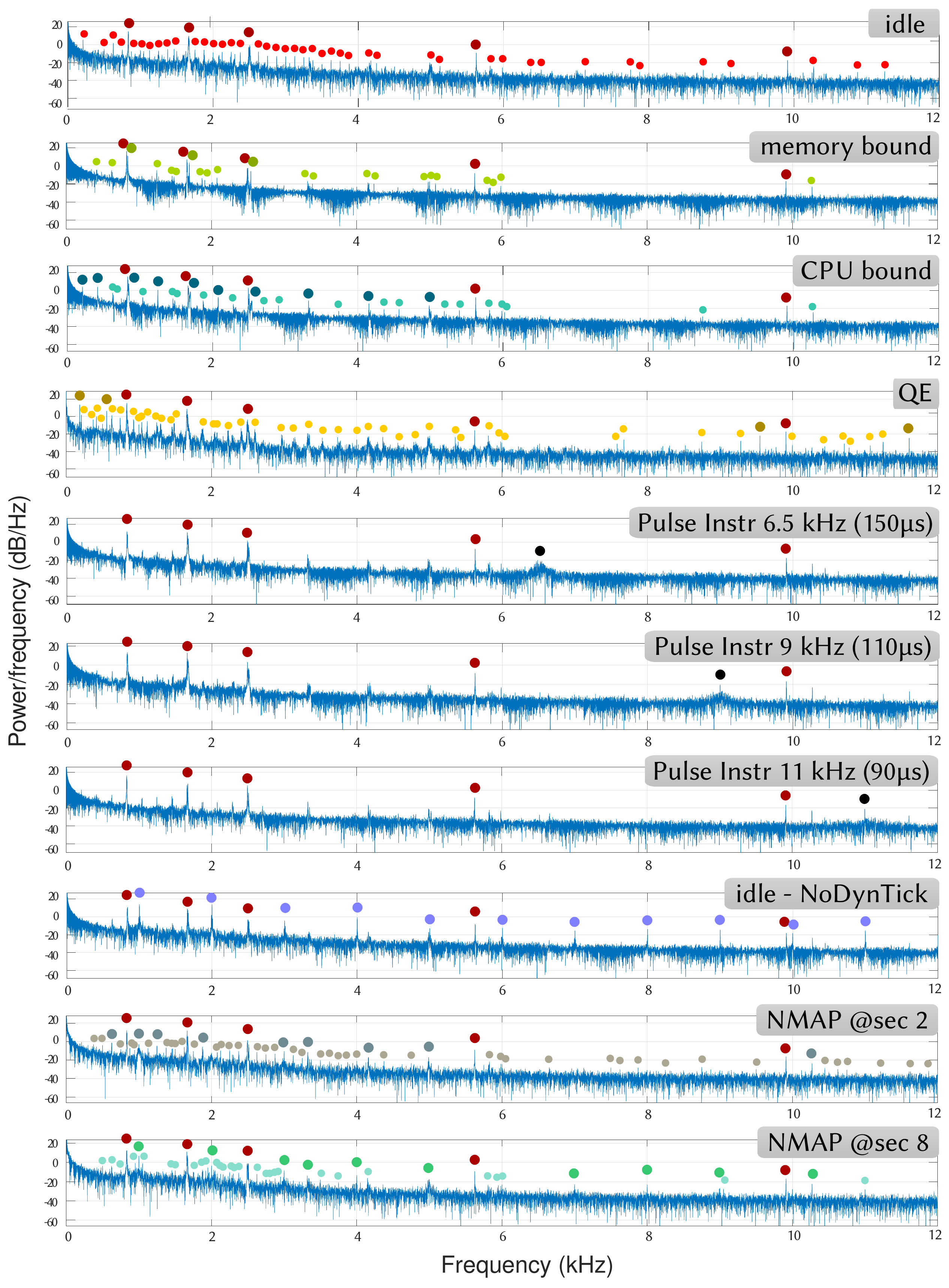}
  \caption{Example of PSD patterns of real bottlenecks and applications.}
  \label{fig:all_FFT}
\end{figure*}

The second set of benchmarks, reported in Figure~\ref{fig:all_FFT}, aims at demonstrating distinctive frequency-domain signatures of real bottlenecks (\textit{e.g.}, CPU or memory limitations) and scientific applications. Goal of this test is not to analyze in depth the reasons behind the peaks, but instead to show that different patterns emerge in the power spectrum with different workloads, which can be used as input features for DNN algorithms. 

In particular, comparing the first four plots we can clearly see four different patterns (peaks highlighted with dark\,/\,light circles to indicate stronger\,/\,weaker magnitude): the first plot portrays the PSD of the computing node in idle and reveals five main peaks (dark red circles) plus other weaker peaks (light red circles) spread in the entire bandwidth \SIrange[range-phrase = --]{0}{12}{\kHz} (richer activity in \SIrange[range-phrase = --]{0}{4}{\kHz}); notice that these main peaks persist also in all other benchmarks; the second and third plot depicts respectively a memory bound and a CPU bound synthetic benchmark, where the former is bound in the SDRAM, while the latter is stuck in the CPU \textit{'front-end'} process (\textit{i.e.}, phase where instructions are fetched and decoded into operations - it differs from the CPU \textit{'back-end'} process where instead the required computation is performed); these two benchmarks report a rich activity up to \SI{6}{\kHz} and are almost flat for frequencies above; moreover, they can be clearly distinguished from their pattern (three main peaks in the tested memory bound application vs. ten main peaks in the CPU bound). It is noteworthy that, while measuring only a power consumption coarse-grain value would not be enough to discern any difference between the two bottlenecks, \DIG (w.r.t SoA monitoring systems) can detect spectral components associated to different usage of architectural resources in the two benchmarks. Finally, the fourth plot shows a real scientific application (\textit{i.e.}, Quantum Espresso - QE~\cite{QE}) and reports four main peaks more with respect to idle (dark yellow circles) and rich activity in all the spectrum.

With the next three plots in Figure~\ref{fig:all_FFT}, we demonstrate that we can appreciate the activity of short regions of code. More in detail, we report the PSD of the computing node when it is running sets of instructions at a desired frequency and duty cycle (\textit{i.e.}, pulse train of instructions where we alternate high load computation phases with idle phases). The black circles in the plots highlight we can capture the activity of software routines running roughly every \SI{150}{\micro\s}, \SI{110}{\micro\s} and \SI{90}{\micro\s}, with a respective duration of \SI{75}{\micro\s}, \SI{55}{\micro\s} and \SI{45}{\micro \s} (\textit{i.e.}, routines at \SI{6.5}{\kilo \Hz}, \SI{9}{\kilo \Hz} and \SI{11}{\kilo \Hz} with \SI{50}{\percent} duty cycle between idle and computation).

Finally, we report in the last three plots of Figure~\ref{fig:all_FFT} two use-cases of anomaly detection in the computing nodes. In particular, the first plot corresponds to a case of misconfiguration, where we disabled in the system the dynamic tick. This is enabled by default in Linux OS in order to potentially make the system more energy efficient (\textit{i.e.}, the kernel can save power when idle because it does not have to wake up regularly just to service the timer tick). Comparing the PSD of the system in idle with the dynamic tick enabled (first plot) with the one without dynamic tick we can clearly see the peak at \SI{1}{\kHz} (frequency of the static tick) and all its harmonics (at multiples of the fundamental) till \SI{11}{\kHz}. We envision this kind of high resolution monitoring along with edge ML analytics to help on catching anomalies in next generation of data centers.

Another interesting scenario for detection of anomalies is related to the detection of cyber-attacks. Network security is a crucial challenge in data centers and cloud infrastructures, to prevent attackers from getting access into the system and steal sensitive data or illegally use computing resources~\cite{minigTahir17,cloudSecTend17}. Before intruding into the system attackers need to gather information about the target machine and its running services, and thus about vulnerabilities that can be exploited. This is called scanning phase, and one of the most popular tools for port scanning is \textit{NMAP}~\cite{nmap_book}.

The use case scenario is an attacker that tries to collect information about the front-end node of data centers\,/\,clouds to get access into the local network. Thus we run NMAP from a remote computer (outside the local network) with the OS detection mode enabled, which means we want to understand open ports, running services and OS of the front-end node. The scanning attack requires around 10 seconds and the last two plots of Figure~\ref{fig:all_FFT} show the PSD of the \DIG power measurements when monitoring the front-end node. With the goal to show that different phases of NMAP correspond to different patterns of PSD, which are also different from the PSD of the system in idle, we report the attacked node at second 2 and second 8, on the first and second plot, respectively. As can be seen, results show 2 different patterns: the plot at second 2, with regards to plot at second 8, reports main peaks only up to \SI{5}{\kilo\Hz} and is almost flat for the frequencies above; instead, the plot at second 8 reports main peaks spread in the entire bandwidth (\textit{i.e.}, $1, 2, 3, 3.5, 4, 7, 8, 9,$ \SI{10.5}{\kHz}). This pattern recognition analysis, based on high resolution power measurements, can be used in future works to exploit ML classifiers running on the edge and correlate this information with SoA signature-based IDS (\textit{e.g.}, SNORT) to help on preventing from intrusions.

\subsection{Edge ML Benchmarking}\label{sub:edge_ML}

In this section we show the performance of ResNets~\cite{ResNet2016} running on the embedded monitoring platform with different layers and sizes on a dataset of pre-computed PSDs of $2048$ points in frequency each (\textit{i.e.}, \SI{4096}{\byte} per spectrogram, which correspond to a time window of \SI{40}{\milli \s}), to show an example of the capability of \DIG on carrying out edge ML inference. In particular, we use TensorFlow inference compiled for ARM architecture in order to exploit the NEON SIMD accelerator.

\begin{figure}[ht]
  \centering
  \includegraphics[width=8.2cm]{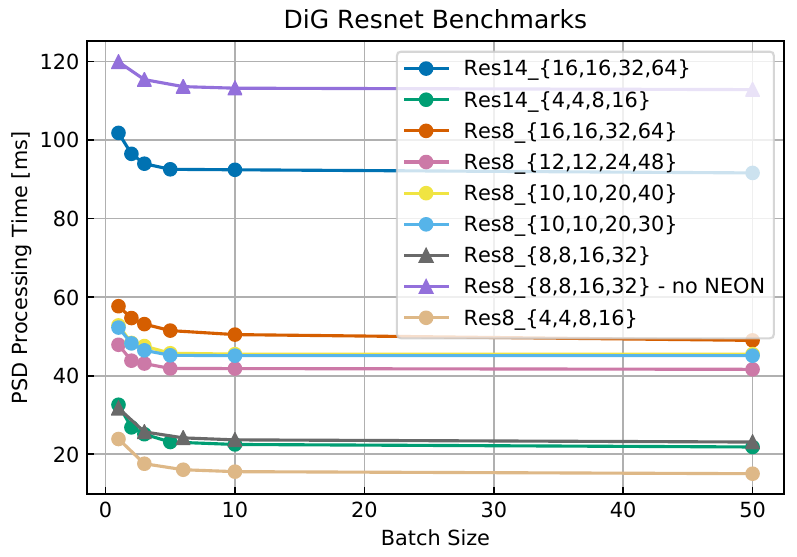}
  \caption{ResNet benchmarking with TensorFlow on BBB.}
  \label{fig:dig_ML}
\end{figure}

Figure~\ref{fig:dig_ML} reports the results of the benchmarks, where the x-axis corresponds to the chosen batch size (\textit{i.e.}, number of test images per iteration) and the y-axis to the processing time per image (milliseconds). Results show that the best trade-off is between batch sizes 3 and 5, which can give an improvement up to around \SI{10}{\milli\s} w.r.t. using the same ResNet with no batch mode enabled. For bigger batch sizes the improvement is negligible (in the best case up to \SI{1}{\milli\s}). Moreover, as a matter of comparison we have run a ResNet with 8 layers and channels $\{8,8,16,32\}$ both exploiting the NEON accelerator (gray triangles) and without using it (purple triangles), and results show a $3.8\times$ improvement when using NEON, which is a relevant processing time when handling high resolution measurements and live analysis.

Finally, considering a time constraint of \SI{40}{\milli\s} for running real-time PSDs on the edge (\SI{4096}{\byte} per spectrogram), results suggest that ResNet with 8 layers and channels $\{8,8,16,32\}$ (and below), but also 14 layers and channels $\{4,4,8,16\}$ (and below), are suitable for running on high resolution power measurements, while larger ResNet size and layers can be used together with performance measurements acquired at a lower rate. It should be noted that in our benchmarks we did not use any optimized ML framework for our embedded platform, such as ARM NN SDK~\cite{ArmNN} or TensorFlow Light~\cite{TF_Light}, which we believe would further improve these results.

\section{Related Work}\label{sec:rw}

Existing off-the-shelf methods to measure power and performance of computing nodes in data centers rely on in-band or out-of-band telemetry depending on the technology vendors. In particular, an example of in-band solution is Intel RAPL~\cite{rapl_Khan2018}, while examples of out-of-band solutions are IBM Amester~\cite{rosedahl2017} and the two standards IPMI~\cite{IPMI} and Redfish~\cite{redfish2017} (\textit{i.e.}, new protocol for managing data centers hardware, that fixes the security vulnerabilities of IPMI~\cite{rosedahl2017}). All these built-in tools allow a fine grain per-component monitoring (\textit{i.e.}, based on hardware performance counters and up to \SI{1}{\s} for IPMI, \SI{1}{\milli\s} for RAPL and \SI{250}{\micro\s} for Amester~\cite{rapl_Khan2018,rosedahl2017,HDEEM}), but not high-resolution power monitoring (\textit{i.e.}, covering the entire signal bandwidth - tens of microseconds).

Pushed by the growing interest in fine-grained power monitoring, industry and academia researchers are providing custom solutions for data centers. Examples are HDEEM~\cite{HDEEM}, PowerInsight~\cite{PowerInsight} and HAEC~\cite{Ilsche2018}. The first two systems provide power consumption measurements up to millisecond time resolution, while the last one has a much more fine grain insight, with a sampling rate up to \SI{500}{\kilo S/\s}. All these custom solutions focus on only monitoring the power consumption (\textit{i.e.}, no performance knowledge) and send all the measurements to a centralized monitoring unit for analysis. Thanks to them, new opportunities for research on energy efficiency and other challenges are now possible, but going toward high resolution measurements this kind of monitoring design entails scalability issues (\textit{e.g.}, as discussed in~\cite{Ilsche2018}, HAEC is suitable for high resolution monitoring in just a node, but not for an entire cluster).

\textit{Comparison with SoA:} In our system (\textit{i.e.}, \DIG) we combined all these features to enable research on several challenges for analytics, automation and control of data centers, with a highly-flexible monitoring platform: (i)~we work completely out-of-band (\textit{i.e.}, no impact\,/\,perturbation on the computing resources); (ii)~we collect all performance counters and (iii)~the full power bandwidth at the plug (\textit{i.e.}, sampling at \SI{50}{\kilo S/\s}) (iv)~with high precision (\textit{i.e.}, below \SI{1}{\percent} - $\sigma$); (v)~we provide highly synchronized measurements (\textit{i.e.}, sub-microsecond) for a detailed correlation of the activities within the cluster; (vi)~we leverage the monitoring between edge and a centralized unit, by exploiting dedicated embedded computers to collect measurements (they have complete knowledge of the status of their node) and have the possibility to carry out both edge and cluster-level analytics; (vii)~the system is scalable (thanks to our flexible design, based on edge monitoring agents and a robust and scalable protocol - MQTT - to the centralized monitoring, where we analyze data at a lower rate), (viii)~technology agnostic (\textit{i.e.}, tested on ARM, Intel and IBM) and (ix)~low cost (\textit{i.e.}, no motherboard redesign required).

\section{Conclusion}\label{sec:end}

This work reports on the design of a novel monitoring infrastructure - namely \DIG~- that enables real-time high-resolution profiling and analytics of data centers, for their automation and control. Main design choices include complete out-of-band monitoring of power and performance, with a dedicated embedded computer per node to perform edge analysis, and a custom power sensor at the plug for high-resolution and high-precision measurements. We report (i)~architecture design choices of both hardware and software, (ii)~monitoring platform performance, (iii)~an extensive set of tests based on Fourier analysis to show the high resolution monitoring insights and (vi)~a campaign of benchmarks of ML inference, running on the embedded computer, to provide an overview of the real-time edge analytics capabilities of \DIG.

\begin{acknowledgements}
This work has been partially supported by the EU H2020 FET project OPRECOMP (g.a. 732631), the EuroLab-4-HPC project, the Italian supercomputing center CINECA and E4 Computer Engineering SpA.
\end{acknowledgements}

\bibliographystyle{spmpsci}      

\end{document}